\documentclass[11pt]{article}
\usepackage{graphicx}
\usepackage{amssymb}
\usepackage{amsmath}
\usepackage{graphicx}
\usepackage{amstext}
\usepackage{esint}

\def\beq{\begin{eqnarray}}    
\def\eeq{\end{eqnarray}}      

\newcommand{\Omo}{\Omega_m^0}

\newcommand{\Oro}{\Omega_{r}^0}

\newcommand{\OLo}{\Omega_{\Lambda}^0}

\newcommand{\rmo}{\rho_{m}^0}

\newcommand{\rmr}{\rho_m}

\newcommand{\rL}{\rho_{\CC}}

\newcommand{\rLo}{\rho_{\CC}^0}

\newcommand{\CC}{\Lambda}

\newcommand{\be}{\begin{equation}}
\newcommand{\ee}{\end{equation}}

\begin{document}



 \hyphenation{cos-mo-lo-gi-cal
sig-ni-fi-cant}




\begin{center}
{\bf\Large Vacuum models with a linear and a quadratic term in H:
structure formation and number counts analysis} \vskip 2mm

 \vskip 8mm

\textbf{Adri\`a G\'omez-Valent, and  Joan Sol\`{a}}

\vskip 0.5cm

High Energy Physics Group, Dept. ECM Univ. de Barcelona,\\
Av. Diagonal 647, E-08028 Barcelona, Catalonia, Spain

\vskip0.15cm and \vskip0.15cm

Institut de Ci{\`e}ncies del Cosmos\\
Univ. de Barcelona, Av. Diagonal 647, E-08028 Barcelona

\vskip0.5cm

\vskip0.4cm

E-mails:    adriagova@ecm.ub.edu, sola@ecm.ub.edu

 \vskip2mm

\end{center}
\vskip 15mm

\begin{quotation}
\noindent {\large\it \underline{Abstract}}. We focus on the class of
cosmological models with a time-evolving vacuum energy density of
the form $\rho_\Lambda(H)=C_0+C_1 H+C_2 H^2$, where $H$ is the
Hubble rate. Higher powers of $H$ could be important for the early
inflationary epoch, but are irrelevant afterwards. We study these
models  at the background level and at the perturbations level, both
at the linear and at the nonlinear regime. We find that those with
$C_0=0$ are seriously hampered, as they are unable to fit
simultaneously the current observational data on Hubble expansion
and the linear growth rate of clustering. This is in contrast to the
$C_0\neq 0$ models, including the concordance $\Lambda$CDM model. We
also compute the redshift distribution of clusters predicted by all
these models, in which the analysis of the nonlinear perturbations
becomes crucial. The outcome is  that the models with $C_0=0$
predict a number of counts with respect to the concordance model
which is much larger, or much smaller, than the $\CC$CDM and the
dynamical models with $C_0\neq 0$. The particular case
$\rho_\Lambda(H)\propto H$ (the pure lineal model), which in the
past was repeatedly motivated by several authors from QCD arguments
applied to cosmology, is also addressed and we assess in detail its
phenomenological status. We conclude that the most favored models
are those with $C_0\neq 0$, and we show how to discriminate them
from the $\CC$CDM.
\end{quotation}
\vskip 5mm

\newpage


\newpage


\section{Introduction}

The accurate measurement of the luminosity-redshift curve of distant
type Ia supernovae carried out at late 1990s by The Supernova
Cosmology Project (Perlmutter et al. 1998) and The High-z Supernova
Search Team (Riess et al. 1998) showed that our universe is speeding
up. This positive acceleration could be produced by the presence of
a  tiny cosmological constant (CC) in Einstein's field equations,
$\CC>0$. This framework, the so-called concordance or $\Lambda$CDM
model, seems to describe quite well the available cosmological data
(Ade et al. 2013). Despite this, the CC, which is usually associated
to the energy density carried by the vacuum, through the parameter
$\rL=\CC/(8\pi\,G)$ (in which $G$ is the Newtonian constant), has
also been the origin of two of the most important current open
problems in physics, namely the old CC problem ({Weinberg 1989) and
the Cosmic Coincidence problem (see e.g. the reviews by Padmanabhan
2003, Peebles \& Ratra 2003, Copeland, Sami \& Tsujikawa 2006). The
severity of these problems are the main motivation to search for
alternative frameworks capable to offer a more satisfactory
explanation for them while still keeping a good fit to the
observational data.

Different scenarios have been proposed in order to alleviate this
situation, to wit: scalar fields, e.g. quintessence, modified
gravity theories, decaying vacuum models, etc (cf. the previous
review articles and references therein). The present work takes the
point of view that the vacuum energy density is a dynamical variable
in QFT in curved spacetime as in such framework it should be
possible to better tackle the basic CC problems\,\footnote{For a
recent review of the idea of dynamical vacuum energy, see  (Sol\`a,
2013) and references therein, and also (Sol\`a \& G\'omez-Valent, 2015) for  additional considerations.}. Our aim here is mainly
phenomenological and hence of eminently practical nature. We extend
the analysis performed in (Basilakos, Plionis \& Sol\`a 2009;
Grande, Sol\`a, Basilakos \& Plionis 2011; G\'omez-Valent, Sol\`a \&
Basilakos 2014), where some dynamical vacuum models based on powers
of the Hubble rate were studied at the background and perturbation
level -- see also (Sol\`a \& Stefancic 2005, 2006).

In this article we focus on the dynamical vacuum models that include
a linear and a quadratic term in $H$, i.e. $\rho_\Lambda(H)=C_0+C_1
H+C_2 H^2$. We discuss the various possibilities, in particular we
examine the phenomenological status of the models where no additive
term $C_0$ is present. Of especial significance is to check out the
purely linear model $\rL\propto H$, which is a particular case of
the $C_0=0$ models. The linear model was amply discussed several
times in the literature by different authors from different points
of view. It was theoretically motivated as a possible fundamental
description of the cosmological vacuum energy in terms of QCD -- see
e.g. (Schutzhold 2002; Klinkhamer \& Volovik 2009; Thomas, Urban \&
Zhitnitsky 2009; Ohta 2011). Phenomenological analysis claiming its
possible interest for the description of the current Universe were
carried out e.g. in  (Borges et al. 2008; Alcaniz et al. 2012;
Chandrachani {\it et al.} 2014). We shall revisit the linear
model here, but only as a particular case of the larger class
of dynamical vacuum models that we analyze. We put to the test all
these models in the light of the recent observational data and
assess which are the most favored ones\,\footnote{A
generalization of the vacuum structure of these models with higher
powers of the Hubble rate, i.e. $H^n\,(n>2)$, has been recently used
to describe inflation in the early universe, see e.g. (Lima,
Basilakos \& Sol\`a 2013, 2014). }.

The layout of this paper is as follows. We address the background
solution of the different models in Section 2. The matter
perturbations (linear and nonlinear) are considered in Section 3.
The confrontation with the linear structure data is performed in
Sect. 4, whereas in section 5 we probe the models with the number
counts method, which requires the nonlinear analysis of structure
formation. In Section 6, we present our conclusions. Finally, in the
Appendix A we briefly extend the discussion of the linear model.


\section{Different types of vacuum models with linear term in H}\label{sect:VacuumModels}
\noindent Let us consider a (spatially) flat FLRW universe. From
Einstein's field equations, i.e. $G_{\mu\nu}=8\pi GT_{\mu\nu}$, one
can derive Friedmann's equation and the pressure equation by taking
the $00$ and the $ij$ component, respectively:
\be\label{eq:FriedmannEq} 3H^2=8\pi G(\rho_\Lambda+\rho_m)\,, \ee
\be\label{eq:PressureEq} 3H^2+2\dot{H}=-8\pi G(p_\Lambda+p_m)\,, \ee
\noindent where the overdot denotes a derivative with respect to the
cosmic time. If we are interested in describing the structure
formation process, we can limit ourselves to consider only the
contributions of cold matter, $p_m=0$, and a true dynamical vacuum
term, $p_\Lambda=-\rho_\Lambda$. Effects of radiation will be
included in a subsequent stage, when they will be necessary.
Combining Eqs.~(\ref{eq:FriedmannEq})~and~(\ref{eq:PressureEq}) it
is easy to obtain the equation of local covariant conservation of
the energy for a pressureless matter fluid:
\be\label{eq:ConservationEq}
\dot{\rho}_m+3H\rho_m=-\dot{\rho}_\Lambda\,. \ee
From these equations we can also obtain the evolution law for the
Hubble function:
\be\label{eq:HubbleEq} \dot{H}+\frac{3}{2}H^2=4\pi
G\rho_\Lambda=\frac{\CC}{2}\, . \ee
\noindent The universe's dynamics depends on the specific dynamical
nature of $\rho_\Lambda$. In the present paper we study the
following dynamical vacuum models\,\footnote{For recent related
studies, see (Basilakos, Plionis \& Sol\`a 2009; Basilakos 2009;
Grande, Sol\`a, Basilakos \& Plionis 2011; and G\'omez-Valent,
Sol\`a \& Basilakos 2014).}:
\begin{eqnarray}\label{eq:Models}
{\rm I}: \phantom{XXX} \rho_\Lambda(H)&=&\frac{3\epsilon H_0}{8\pi G} H\nonumber\\
{\rm II}: \phantom{XXX}\rho_\Lambda(H)&=&\frac{3}{8\pi G}(\epsilon H_0H+\nu H^2)\\
{\rm III}: \phantom{XXX}\rho_\Lambda(H)&=&\frac{3}{8\pi G}(c_0+\epsilon H_0H)\nonumber\\
{\rm IV}:\phantom{XXX} \rho_\Lambda(H)&=&\frac{3}{8\pi
G}(c_0+\epsilon H_0H+\nu H^2)\,. \nonumber
\end{eqnarray}
Notice that the parameter $c_0\equiv\left(8\pi\,G/3\right)\,C_0$ has
dimension $2$ (i.e. mass squared) in natural units. We have
introduced the dimensionful constant $H_0$  (the value of the Hubble
function at present) as a part of the linear term, and in this way
the parameter $\epsilon$ in front of it can be dimensionless.
Similarly $\nu$ is dimensionless since it is the coefficient of
$H^2$. Obviously models I, II and III are particular cases of IV,
i.e. they can be obtained from IV just by setting $c_0=\nu=0$,
$c_0=0$ and $\nu=0$, respectively. However, it is convenient to
study the different implementations separately because they are not
phenomenological alike, and some of them are actually unfavored.

On inspecting the structure of these models, it is even questionable
that theoretically all these possibilities are admissible. For
example, the presence of a linear term $\propto H$ in the structure
of $\rL(H)$ deserves some considerations. This term does not respect
the general covariance of the effective action of QFT in curved
spacetime (Shapiro \& Sol\`a, 2009). The reason is that it involves
only one time derivative with respect to the scale factor. In
contrast, the quadratic terms $\propto H^2$ involve two derivatives
and hence it can be consistent with covariance. From this point of
view one expects that the term $\propto H^2$ is a primary structure
in a dynamical $\rL$ model, whereas $\propto H$ is not. Still, we
cannot exclude a priori the presence of the linear terms since they
can be of phenomenological interest. For example, they could mimic
bulk viscosity effects (cf. Barrow 1983; Zimdahl 1996; Ren \& Meng
2006; Komatsu \& Kimura 2013).

The first task to do in order to analyze the above models is to
solve the background cosmological equations. In the following we
provide the solution of model IV, which is the more general one.
However, in this model the Hubble function and the energy densities
cannot be solved explicitly in terms of the scale factor, only in
terms of the cosmic time. This feature also holds for model III.
Models I and II, however, can be solved analytically also as a
function of the scale factor and we will do it because it is more
convenient.

For all these models we have the following relation between the
basic parameters $c_0$, $\nu$ and $\epsilon$, which follows from
imposing that the dynamical vacuum energy density $\rL(H)$ must
coincide with the current value $\rLo$ at present $t_0$ (the point
where $H(t_0)=H_0$):
\begin{equation}\label{eq:C0TypeB}
c_0=\frac{8\pi
G}{3}\rho^0_\Lambda-H_0^2(\epsilon+\nu)=H_0^2(\OLo-\epsilon-\nu)\,.
\end{equation}
\noindent If we apply Eq.\,(\ref{eq:HubbleEq}) to the general
type-IV models, we find
\be\label{eq:diffEq.TypeB} \frac{2}{3}\dot{H}+\zeta\,H^2-\epsilon
H_0 H=c_0\, , \ee
\noindent where we have defined $\zeta\equiv 1-\nu$. Upon direct
integration we obtain the Hubble function as a hyperbolic function
of the cosmic time:
\be \label{eq:Hubblef}
H(t)=\frac{H_0}{2\,\zeta}\left[\mathcal{F}\,\coth\left(\frac{3}{4}H_0\mathcal{F}\,t\right)+\epsilon\right]\,,
\ee
\noindent with
\be\label{eq:defmathF}
\mathcal{F}(\OLo,\epsilon,\nu)\equiv\sqrt{\epsilon^2+4\,\zeta(\OLo-\epsilon-\nu)}\,.
\ee
\noindent Using (\ref{eq:Hubblef}) in the expression of
$\rho_\Lambda(H)$ for type-IV models we can infer the vacuum energy
density in terms of $t$
\be\label{eq:rhoL} \rho_\Lambda(t)=\frac{3H_0^2}{32\pi
G\zeta^2}\left[\mathcal{F}^2+\epsilon^2+2\epsilon\mathcal{F}\coth\left(\frac{3}{4}H_0\mathcal{F}t\right)+\nu\mathcal{F}^2\,{\rm
csch}^2\left(\frac{3}{4}H_0\mathcal{F}t\right)\right]\,. \ee
Inserting equations (\ref{eq:Hubblef}) and (\ref{eq:rhoL}) into
Friedmann's equation (\ref{eq:FriedmannEq}) we can also derive the
time evolution of the pressureless matter density:
\be\label{eq:rhom} \rho_m(t)=-\frac{\dot{H}(t)}{4\pi
G}=\frac{3H_0^2}{32\pi G\,\zeta}\mathcal{F}^2\,{\rm
csch}^2\left(\frac{3}{4}H_0\mathcal{F}t\right)\,. \ee
The scale factor $a(t)$ can be obtained by integration of
Eq.\,(\ref{eq:Hubblef}):

\be\label{eq:atBtype} a(t)=B^{-\frac{1}{3\zeta}}e^{\frac{\epsilon
H_0}{2\,\zeta}t}\,\sinh^{\frac{2}{3\zeta}}\left(\frac{3}{4}H_0\mathcal{F}t\right)\,,
\ee
with the normalization constant ($a(t_0)=1$):
\begin{equation}\label{eq:defCtypeBomega}
B=\left[\frac{[(2\zeta-\epsilon)^2-\mathcal{F}^2]^{1+\frac{\epsilon}{\mathcal{F}}}}{\mathcal{F}^2(\mathcal{F}+2\zeta-\epsilon)^{\frac{2\epsilon}{\mathcal{F}}}}\right]^{-1}\,.
\end{equation}
\noindent From (\ref{eq:atBtype}) one sees that, in general, it is
not possible to eliminate the cosmic time in terms of the scale
factor. It is only possible if $\epsilon=0$ and/or $c_0=0$.

Let us remark that for models III and IV the values of $\epsilon$
and $\nu$ should necessarily be small since they parametrize a mild
dynamical departure from the $\CC$CDM which we know it fits
reasonably well the data. The dynamical vacuum function $\rL(H)$ in
these models stays around the constant value $\rLo$ for $H$ near
$H_0$ and hence $\epsilon$ and $\nu$ must be small in absolute
value. This situation is of course possible because for these models
$c_0\neq0$. Put another way: models III and IV have a smooth
$\CC$CDM limit for $c_0\to 0$, in contrast to models I and II.  As
we shall see in Sect. \ref{sect:LinearGrowth}, the confrontation of
models III and IV against observations does indeed confirm that
$|\epsilon|$ and $|\nu|$ are in the ballpark of $\sim 10^{-3}$ (see
Table \ref{tableFit}). Quite in contrast, for models I and II
$\epsilon$ and $\nu$ cannot be arbitrarily small since $\rL(H)$ for
these models is not protected by the nonvanishing additive term
$c_0$. For them, the constraint (\ref{eq:C0TypeB}) implies that the
following relations must hold:
\be {\rm I}:\quad\epsilon=\Omega^0_\Lambda\quad (\nu=0)\quad \ \ \ \
{\rm
II}:\quad\epsilon+\nu=\Omega^0_\Lambda\,.\label{eq:constraintsIandII}
\ee
It is thus clear that in model I there is no free parameter (apart
from $\Omo$ or $\OLo$), and for model II we find that if $\epsilon$
is small, $\nu$ cannot be small, and vice versa.

Obviously models I and II satisfy one of the aforementioned
conditions for which the solution in terms of the scale factor is
possible, so let us provide such analytical solution in this case.
The constraints (\ref{eq:constraintsIandII}) entail that the
quantity $\mathcal{F}$ defined in (\ref{eq:defmathF}) boils down to
$\mathcal{F}\to\epsilon$, and this allows to combine the exponential
factor and the hyperbolic function in (\ref{eq:atBtype}). The result
for model II reads:
\be \label{eq:atC1}
a(t)=\left(\frac{\Omo}{\zeta-\Omo}\right)^{2/3\zeta}\left[
e^{3\,(\zeta-\Omo) H_0\,t/2}-1 \right]^{2/3\zeta}\,. \ee
\noindent From here we can invert and derive $t(a)$, and then
substitute in (\ref{eq:Hubblef}) to obtain the normalized Hubble
rate to its current value, i.e. $E(a)\equiv H(a)/H_0$, for type-II
models:
%
\be\label{eq:E2aC1} E(a)=1+\frac{\Omo}{\zeta}
\left(a^{-3\zeta/2}-1\right)\,. \ee
\noindent We can also furnish analytical expressions for the matter
and vacuum energy densities as a function of the scale factor:
\be\label{eq:rhomaC1}
\rho_m(a)=\rho^0_c\left[\zeta\,E^2(a)-(\zeta-\Omo)
E(a)\right]=\rho_m^0\,f(a)\,a^{-3\zeta}\,, \ee
where
\be\label{eq:fa}
f(a)=a^{3\zeta/2}\,E(a)=\frac{\Omo}{\zeta}+\left(1-\frac{\Omo}{\zeta}\right)\,a^{3\zeta/2}\,,
\ee \noindent and
\be\label{eq:rhoLaC1}
\rho_\Lambda(a)=\rho^0_c\left[(1-\zeta)\,E^2(a)+(\zeta-\Omo)
E(a)\right]\,. \ee
Here $\rho^0_c$ is the current critical density, i.e.
$\rho^0_c=3H_0^2/8\pi G$. The corresponding expressions for type-I
model can be directly extracted from (\ref{eq:E2aC1}),
(\ref{eq:rhomaC1}) and (\ref{eq:rhoLaC1} by setting $\zeta=1$):
\begin{eqnarray}\label{eq:rhomaLinear1}
H(a)=H_0\left[1+\Omo \left(a^{-3/2}-1\right)\right]\,,\phantom{XXX}\\
\rho_m(a)=\rho_m^0\left[\Omo+(1-\Omo)a^{3/2}\right]\,a^{-3}\label{eq:rhomaLinear2}\,,\phantom{X}\\
\rho_\Lambda(a)=\rho^0_\Lambda\frac{H(a)}{H_0}=\rho^0_\Lambda\left[1+\Omo
\left(a^{-3/2}-1\right)\right]\label{eq:rhoLaLinear2}\,.
\end{eqnarray}
\noindent The first term on the {\it r.h.s.} of
Eq.~(\ref{eq:rhomaC1}) is the dominant one at high redshifts ($z\gg
1$, equivalently $a\ll 1$). Therefore, in this regime the matter
density evolves like $\rho_m\propto \Omega^{(0)2}_m
a^{-3\zeta}\propto \Omega^{(0)2}_m(1+z)^{3\zeta}$. A similar scaling
law is found for type-I models, with $\zeta=1$.

The following observation is in order. In the concordance model we
have the standard behavior of the matter density
$\rho_m(z)=\rho_m^0\,(1+z)^{3}$, but when we compare it with
(\ref{eq:rhomaC1}) and (\ref{eq:rhomaLinear2}) we observe that for
type-I and II models there is in an extra factor of $\Omo$. This
factor stands out maximally in the remote past where  for the same
value of $\rho_m$ the type-I and type-II models should predict a
larger matter density $\Omo$ at present (cf. Appendix A). The reason
is obvious: if a term of order $\left(\Omo\right)^2$ should mimic
the standard value $\left(\Omo\right)^{\CC{\rm CDM}}\simeq0.3$, the
value itself of $\Omo$ must be of order $\sim 0.5$ and hence
significantly larger ($\sim70\%$) than the standard one. This
situation does not occur so acutely for the low and intermediate
redshift range, as can be seen e.g. from
Eq.\,(\ref{eq:rhomaLinear2}) for model I, where for $a\simeq 1$ the
two terms in the square brackets add up approximately to $1$ and we
recover the $\CC$CDM behavior $\rho_m\sim \rmo\,a^{-3}$. As we will
comment in Sect. \ref{sect:LinearGrowth}, we find more appropriate
to test these ``anomalous'' models near the region where they can
mimic the $\CC$CDM to some reasonable extent, i.e. at relatively low
redshifts.

Up to now, we have not included the effect of relativistic matter
since we were interested in studying the background solutions near
our time and the physics of cosmological perturbations. In spite of
this, when we will put our models to the test the radiation
correction must be taken into account in our overall fit to the main
cosmological data, especially in regard to the data on Baryonic
Acoustic Oscillations (BAOs) and Cosmic Microwave Background (CMB).
In fact, when the CMB was released ($z_{*}\sim 1100$) the amount of
radiation was not negligible, so we had better include the
relativistic matter component in our analysis. The generalized
energy conservation law involving also radiation reads as follows:
\be \dot{\rho}_m+\dot{\rho}_r+3H\rho_m+4H\rho_r=-\dot{\rho}_\Lambda
\, . \ee
\noindent We may compute $\dot{\rho}_\Lambda$ in this expression
from the explicit form of the general vacuum energy in
Eq.\,(\ref{eq:Models}), i.e. using model IV. Since relativistic and
non-relativistic matter are in interaction one can split the
obtained expression with the aid of an interaction source $Q(t)$:
\begin{eqnarray}\label{eq:splitQB}
\phantom{XXXXX}\dot{\rho}_m+3\rho_m \left[\zeta H -\frac{\epsilon}{2} H_0\right]&=&Q(t)\,,\nonumber\\
\phantom{XXXXX}\dot{\rho}_r+4\rho_r\left[\zeta H -\frac{\epsilon}{2}
H_0\right]&=&-Q(t)\,.
\end{eqnarray}
\noindent Notice that when one of the matter components dominates
over the other we are allowed to turn the source $Q(t)$ off and
solve the decoupled system. The corresponding results for
non-relativistic and relativistic matter are as follows:
\begin{equation}\label{eq:rhomNR}
\rho_m(t,a)=\rho^0_m\,e^{\frac{3}{2}\epsilon
H_0(t-t_0)}\,a^{-3\zeta}\,,
\end{equation}
\begin{equation}\label{eq:rhorNR}
\rho_r(t,a)=\rho^0_r\,e^{2\epsilon H_0(t-t_0)}\,a^{-4\zeta}\,.
\end{equation}
The presence of the time dependence in the exponential, which is
triggered by the $\epsilon$-parameter of the linear term in the
vacuum function, is reminiscent of the fact that for models III and
IV the energy densities cannot be expressed fully in terms of the
scale factor.

In good approximation we can assume that the evolution of the scale
factor as of the time when the CMB was released corresponds to the
cold matter epoch, i.e. we suppose it is evolving as indicated in
Eq.\,(\ref{eq:atBtype}). In this way we can determine the energy
densities (\ref{eq:rhomNR}) and (\ref{eq:rhorNR}) in terms of the
cosmic time only. This last step can be performed analytically only
if we suppose that the effects of radiation are sufficiently small,
as it is indeed the case under consideration. We find:
\begin{equation}\label{eq:rhomNR2}
\rho_m(t)=\rho^0_m\,B\,e^{-\frac{3}{2}\epsilon H_0 t_0}\,{\rm
csch}^2\left(\frac{3}{4}H_0\mathcal{F}t\right)\,,
\end{equation}
\begin{equation}\label{eq:rhorNR2}
\rho_r(t)=\rho^0_r\,B^{4/3}\,e^{-2\epsilon H_0 t_0}\,{\rm
csch}^{8/3}\left(\frac{3}{4}H_0\mathcal{F}t\right)\,,
\end{equation}
where the constant $B$ is the same as that in
(\ref{eq:defCtypeBomega}).

\noindent The following normalization condition must be fulfilled so
that the energy densities take the present value at $t=t_0$:
\begin{equation}\label{eq:condition}
B\,e^{-\frac{3}{2}\epsilon H_0 t_0}\,{\rm
csch}^2\left(\frac{3}{4}H_0\mathcal{F}t_0\right)=1\,.
\end{equation}
Note that we can actually determine $t_0$ in terms of the remaining
parameters by matching equations (\ref{eq:rhomNR2}) and
(\ref{eq:rhom}):
\begin{equation}\label{eq:toBF}
B\,e^{-\frac{3}{2}\epsilon H_0
t_0}=\frac{\mathcal{F}^2}{4\zeta\,\Omo}=\frac{\epsilon^2+4\,\zeta(\OLo-\epsilon-\nu)}{4\zeta\,\Omo}\,.
\end{equation}
\noindent The Hubble function of the matter-dominated epoch
including the radiation contribution can be calculated from the
generalized Friedmann's equation for model IV:
\begin{equation}\label{eq:HtypeBRad}
H^2(t)=\frac{8\pi G}{3}\left[\rho_m(t)+\rho_r(t)\right]+c_0+\epsilon
H_0H(t)+\nu H^2(t)\,.
\end{equation}
\noindent It can be checked that thanks to the condition
(\ref{eq:condition}) the implicit formula (\ref{eq:HtypeBRad}) for
the Hubble function leads to the extended cosmic sum rule
$\Omo+\Omega_r^0+\OLo=1$, as expected. After some rearrangement and
making use of (\ref{eq:rhomNR2}) and (\ref{eq:rhorNR2}), we can
bring Eq.\,(\ref{eq:HtypeBRad}) into the form
\begin{equation}\label{eq:solvingHalgebraic}
\zeta H^2-\epsilon H_0 H- H_0^2\,s(t)=0,
\end{equation}
\noindent where
\be\label{eq:stfunction}
s(t)=\zeta-\epsilon-\Omo-\Oro+\frac{\mathcal{F}^2}{4\zeta}\,{\rm
csch}^2\left(\frac{3}{4}H_0\mathcal{F}t\right)
+\Oro\,\left(\frac{\mathcal{F}^2}{4\zeta\,\Omo}\right)^{4/3}\,{\rm
csch}^{8/3}\left(\frac{3}{4}H_0\mathcal{F}t\right)\,. \ee
\noindent Thus, the sought-for Hubble function in the presence of a
relatively small amount of relativistic matter can be computed by
solving Eq. (\ref{eq:solvingHalgebraic}):
\begin{table*}
\begin{center}
\begin{tabular}{| c |  c |c | c | c | }
\multicolumn{1}{c}{Model} & \multicolumn{1}{c}{$\Omo$} &
\multicolumn{1}{c}{$\nu=1-\zeta$} & \multicolumn{1}{c}{$\epsilon$} &
\multicolumn{1}{c}{$\chi^2/dof$}  \\\hline $\CC$CDM & $0.293\pm
0.013$ & - & - & $567.8/586$ \\\hline I & $0.302^{+0.010}_{-0.009} $
& -  & $1-\Omo$ & $575.7/585$ \\\hline II & $0.295^{+0.016}_{-0.011}
$ &  $1-\Omo-\epsilon$  & $0.93^{+0.01}_{-0.02}$ & $567.7/584$
\\\hline III & $ 0.297^{+0.015}_{-0.014}$ & -  &
$-0.014^{+0.016}_{-0.013}$ & $587.2/585$  \\\hline ${\rm IVa}$ & $
0.300^{+0.017}_{-0.003}$ &$ - 0.004\pm 0.002 $ & $- 0.004\pm 0.002$
& $583.1/585$ \\ \hline ${\rm IVb}$ & $ 0.297^{+0.005}_{-0.015}$ &$
- 0.002\pm 0.002 $ & $-0.001\pm0.001$ & $579.5/585$ \\ \hline
 \end{tabular}
\caption{\scriptsize The fit values for the various models, together with their
statistical significance according to a $\chi^2$-test. We have
performed a joint statistical analysis of the SNIa+CMB+BAO$_{dz}$
data for the $\CC$CDM, type-III and type-IV models. For type-I and
type-II models, instead, we have used SNIa+BAO$_{A}$ data for the
reasons explained in the text. To break parameter degeneracies we
present the fitting results for two different cases: the one
indicated as IVa (resp. IVb) corresponds to $\nu=\epsilon$ (resp.
$\nu=2\epsilon$). Recall that because of the constraints
(\ref{eq:constraintsIandII}) model I has $\Omo$ as the sole free
parameter, whereas for model II one can adopt $\Omo$ and $\epsilon$.
\label{tableFit}}
\end{center}
\end{table*}

$$H(t)=\frac{H_0}{2\zeta}\,\left[\epsilon +\sqrt{\epsilon^2+4\zeta
s(t)}\right]$$
\begin{equation}\label{eq:hubbleimpro}
H(t)=\frac{H_0}{2\zeta}\,\left[\epsilon
+\mathcal{F}\,\coth{\left(\frac{3}{4}H_0\mathcal{F}t\right)\,\sqrt{1+\Delta(t)}}\right]\,,
\end{equation}
\noindent where $\Delta (t)$ is defined as
\begin{equation}\label{eq:Delta}
\Delta(t)=\frac{\Oro}{\Omo}\,\left(\frac{\mathcal{F}^2}{4\zeta\Omo}\right)^{1/3}\,{\rm
csch}^{2/3}\left(\frac{3}{4}H_0\mathcal{F}t\right){\rm
sch}^2\left(\frac{3}{4}H_0\mathcal{F}t\right)\,.
\end{equation}
\noindent In Eq. (\ref{eq:hubbleimpro}) we have made use of the
extended cosmic sum rule mentioned above to establish the relation
\begin{equation}\label{eq:identity1}
\zeta-\epsilon-\Omo-\Oro=\OLo-\nu-\epsilon=\frac{\mathcal{F}^2-\epsilon^2}{4\zeta}\,.
\end{equation}
\noindent The numerical integration of (\ref{eq:hubbleimpro})
provides the improved form of the scale factor for a general type-IV
model, namely $a(t)=e^{\int_{t_0}^{t}H(\hat{t})d\hat{t}}$. Thus, we
can obtain the points of the curves $H(a)$, $\rho_m(a)$, $\rho_r(a)$
and $\rho_\Lambda(a)$ computationally using the results presented
before. With this strategy we can better confront the model with
observations since the data inputs are given in terms of the
cosmological redshift variable $z=(1-a)/a$.

On comparing equations (\ref{eq:Hubblef}) with
(\ref{eq:hubbleimpro}) we immediately recognize that $\Delta (t)$
represents the correction term introduced by the effect of the
radiation upon the original expression (\ref{eq:Hubblef}). Obviously
$\Delta(t)\approx 0$ at the present time. However, this is not so at
the decoupling time. Indeed, taking\,\footnote{We include photons
and $N_{\nu}=3$ neutrino species, with  $\Omo h^2\simeq 0.14$ from
Planck+WP\,(Ade et al. 2013).}
$\Oro/\Omo=\left(1+0.227\,N_{\nu}\right)\,\Omega_{\gamma}^0/\Omo=4.15\times
10^{-5}\left(\Omo\,h^2\right)^{-1}\simeq 3\times 10^{-4}$, we find
that $\Delta(t)$ rockets into a numerical value of order $\sim 10^3$
at the time of last scattering. The net outcome is that the fraction
of relativistic matter at decoupling can be around 23 $\%$.

As we have seen, for type-II model (and in particular for type-I)
one can derive the energy densities (\ref{eq:rhomNR}) and
(\ref{eq:rhorNR}) in terms of the scale factor. The answer for
nonrelativistic matter is given in Eq.\,(\ref{eq:rhomaC1}). For
radiation we can proceed in a similar way, and the result is:
\be\label{eq:rhorC1} \rho_r(a)=\rho_r^0\,f^{4/3}(a)\,a^{-4\zeta}\,,
\ee
where $f(a)$ was defined in (\ref{eq:fa}). The modified Hubble rate
for type-II models can be expressed in terms of the scale factor by
solving (\ref{eq:HtypeBRad}) after setting $c_0=0$ and using
(\ref{eq:rhomaC1}) and (\ref{eq:rhorC1}):
\be\label{eq:EaIIrhomrhoR}
E(a)=\frac{\zeta-\Omo+\sqrt{(\zeta-\Omo)^2+4\zeta\left[\frac{\rho_m(a)+\rho_r(a)}{\rho_c^{0}}\right]}}{2\zeta}\,.
\ee
Substituting this expression in (\ref{eq:Models})/ type II we find
the corresponding vacuum energy density $\rho_\CC(a)$ including the
effect of radiation, which is a cumbersome expression.

We can also estimate the equality time,  $t_{eq}$, between the
radiation and the non-relativistic matter energy densities  for
models of type I and II. Equating (\ref{eq:rhomaC1}) and
(\ref{eq:rhorC1}) and taking into account that $a_{eq}\ll 1$ we
obtain:
\be\label{eq:aeq}
a_{eq}=\left[\frac{\Omega^{0}_r}{\zeta^{1/3}\left(\Omo\right)^{2/3}}\right]^{1/\zeta}\,.
\ee
For the typical values that $\zeta$ and $\Omo$ take in Table
\ref{tableFit}, $a_{eq}$ deviates significantly from the $\CC$CDM
prediction value $a_{eq}=\Omega^0_r/\Omo$. In contrast, for models
of type-III and IV (which have $c_0\ne 0$) one can use the
concordance value as a very good approximation. In these cases one
can show that
$a_{eq}=\Omega^0_r/\Omo[1+x\,\ln(\Omega^0_r/\Omo)+\mathcal{O}(x^2)]$,
where $x(\epsilon,\nu)\ll 1$ and the deviations from the $\CC$CDM
model value are only at the few percent level. Needless to say,
additional important differences of the $c_0=0$ models are expected
to appear in connection to the photon decoupling and baryon drag
epochs, and in the value of the comoving Hubble scale,
$k^{-1}_{eq}$, at the redshift of matter-radiation equality (cf.
sections \ref{sect:LinearGrowth} and \ref{sect:NumberCounts} for
additional considerations on these matters).

As to the behavior of the energy densities deep in the radiation
epoch for type-I and II models, let us note that it can be relevant
for the primordial big bang nucleosynthesis (BBN). The  ratio between
the vacuum and radiation energy densities for $a\ll 1$ can be
estimated from the foregoing analysis, with the result:
\begin{equation}\label{eq:ratioDens}
\frac{\rho_\CC}{\rho_r}\approx
\left(\frac{1-\zeta}{\zeta}\right)\left[1+\left(\frac{a}{a_{eq}}\right)^\zeta\right]\,,
\end{equation}
\noindent where use has been made of (\ref{eq:aeq}).  Notice that
the term enclosed in the square brackets provides the correction to
the result that can be inferred from (\ref{eq:rhomaC1}) and
(\ref{eq:rhoLaC1}) in the matter dominated epoch at high redshift,
but without radiation. Taking into account the fitted values of the
parameters presented in Table \ref{tableFit}, we find
$\zeta=\epsilon+\Omo=1.225$. Therefore, at $a=a_{eq}$ the ratio
(\ref{eq:ratioDens}) yields $\rho_\CC\approx-0.37\rho_m$, and at the
BBN epoch (where $a\ll a_{eq}$) we have
$\rho_\CC\approx-0.18\rho_m$. We learn from these estimates that
type-II models predict a negative value of $\rho_\CC$ in the past
and, moreover, it is  a non-negligible faction of $\rho_r$ at the
BBN time. This fraction could of course be made smaller by
decreasing $\nu$ (i.e. approaching $\zeta\to 1$) but this would
worsen the quality of the fit since model II provides a better fit
to low energy data than model I (cf. Table I). In compensation for
its poorer description of the current data, model I satisfies
$\rho_\Lambda/\rho_r\to 0$ when $a\to 0$, similar to the $\CC$CDM,
and therefore its vacuum energy is, in principle, harmless for the
BBN.


\section{Linear and nonlinear structure formation}
\noindent For type-IV models the nonlinear equation for the growth
factor $\delta_m(t)=\delta\rho_m(t)/\rho_m(t)$ can be derived after
some lengthy calculations leading to the following final
result\,\footnote{We follow the procedure explained in detail in
Appendix A of (Grande, Sol\`a, Basilakos \& Plionis 2011).}:
$$\frac{9}{16}H_0^2\mathcal{F}^2(1-y)^2\delta_m^{\prime\prime}+\frac{3}{4}H_0\mathcal{F}\delta_m^\prime(1-y^2)\left[2H+\Psi-\frac{3}{2}yH_0\mathcal{F}\right]$$
$$+\left[2H\Psi+\frac{3}{4}H_0\mathcal{F}(1-y^2)\Psi^\prime-\frac{\rho_m}{2}(1+\delta_m)\right]\delta_m-\frac{\Psi^2\delta_m^2}{3(1+\delta_m)}$$
\begin{equation}\label{eq:SDE2}
-\left[\frac{4\left(\frac{3}{4}H_0\mathcal{F}(1-y^2)\delta_m^\prime\right)^2+
\frac{15}{4}\Psi
H_0\mathcal{F}(1-y^2)\delta_m\delta^\prime_m}{3(1+\delta_m)}\right]=0\,,
\end{equation}
where the variable $y$ is related to the cosmic time through
$y(t)\equiv\coth\left(\frac{3}{4}H_0\mathcal{F}t\right)$, and
$\Psi\equiv -\dot{\rho}_\Lambda/\rho_m$. The primes indicate
derivatives with respect to $y$. The expressions for $\Psi(y)$,
$\rho_m(y)$ and $H(y)$ for type-IV models are, respectively:
\begin{equation}\label{eq:Qdot2}
\Psi(y)=-\frac{\dot{\rho}_{\Lambda}(t)}{\rho_m(t)}=
\frac{3H_0}{2\zeta}\,\left[y(1-\zeta)\,\mathcal{F}+\epsilon\right]\,,
\end{equation}
\begin{equation}\label{eq:rhoy}
\rho_m(y)=\frac{3H_0^2}{32\pi G\zeta}\mathcal{F}^2(y^2-1)\,,
\end{equation}
and \be\label{eq:Hubbley}
H(y)=\frac{H_0}{2\zeta}[\mathcal{F}y+\epsilon]. \ee
\noindent The numerical solution of the above nonlinear equation is
used to compute the collapse density threshold $\delta_c(z)$, an
important model-dependent quantity that is used in the number counts
analysis of Sect.\ref{sect:NumberCounts}.  Once more we refer the
reader to (Grande, Sol\`a, Basilakos \& Plionis 2011) for details
(see also Pace, Waizmann \& Bartelmann 2010). \noindent

If we are, however, interested only in the linear growth factor we
can throw away the nonlinear terms from (\ref{eq:SDE2}), i.e. the
${\cal O}(\delta_m^2)$ terms. Let us dispense with these terms at
this point, as we wish to focus on the large scale linear
perturbations. In practice, to solve the resulting linear
differential equation we have to fix the initial conditions for
$\delta_m$ and $\delta_m^\prime$.
We take them at very high redshift $z\gg 1$. The
scale factor (\ref{eq:atBtype}) can be expressed in terms of $y$:
\begin{equation}
a(y)=B^{-\frac{1}{3\zeta}}(y^2-1)^{-\frac{1}{3\zeta}}\left(\frac{y+1}{y-1}\right)^{\frac{\epsilon}{3\zeta\mathcal{F}}}\,.
\end{equation}
\noindent For the general model IV (with $c_0\neq 0$) we normalize the growth factor with the value
$\delta_m(z=0)$, i.e. $\delta_m(a=1)$, and we take $\delta_m(a)=a$
at very high redshifts. The initial conditions at $y_i=700$,
corresponding to $z_i\simeq 100$ for type-IV models, are the
following. For the growth factor we have $\delta_m(y_i)=a(y_i)$, and
for its derivative with respect to the $y$-variable, we obtain
\begin{equation}
\delta^\prime_m(y_i)=\left.\frac{da(y)}{dy}\right|_{y_i}=\frac{-2a(y_i)}{3\zeta(y_i^2-1)}\left(y_i+\frac{\epsilon}{\mathcal{F}}\right)\,.
\end{equation}
Unfortunately, the differential equation for $\delta_m$ cannot be
solved analytically neither for type-III nor for type-IV models. We
are forced to use numerical techniques, for instance the method of
finite differences, which is anyway necessary for tackling the
original nonlinear equation (\ref{eq:SDE2}). For type-II models the
perturbation equations can be readily obtained by setting
$c_0\to 0$. In this limit, the
$y$-variable reads $y=-1+2\zeta E/\epsilon$. Introducing now
$
y_1=(y+1)/(y-1)\,,
$
the differential equation for the linear perturbations becomes

\be\label{eq:perturbC1}
3\zeta^2\,y_1\,(y_1-1)^2\,\frac{d^2\delta_m}{dy_1^2}+2\,\zeta\,(y_1-1)\,(5y_1-3\zeta)\,\frac{d\delta_m}{dy_1}-2\,(2-\zeta)\,(3\zeta-2y_1)\,\delta_m=0\,.
\ee
This result is consistent with that of (Basilakos \& Sol\`a 2014). A
power-like solution of Eq.\,(\ref{eq:perturbC1}) immediately ensues:
$\delta_{m-}(y_1)\sim (y_1-1)^{(\zeta-2)/\zeta}$. While an explicit
relation of the variable $y$ with the scale factor is impossible for
models III and IV, for type-II models the variable $y_1$ defined above
permits such relation:
\begin{equation}
y_1=\frac{\zeta E}{\zeta
E-\epsilon}=1+\frac{\zeta-\Omo}{\Omo}\,a^{3\zeta/2} \,.
\end{equation}
\noindent Thanks to this feature the previously found solution can
be rewritten as $\delta_{m-}(a)\sim a^{3(\zeta-2)/2}$. The latter is
the decaying mode solution (since $\zeta<2 $) and, therefore, must
be rejected. From it we can generate the growing mode solution for
the type-II model:
\be \label{eqff2} \delta_{m+}(a)=C_1\, a^{3(\zeta-2)/2}\int_{0}^{a}
\frac{{\rm d}a'}{a'^{3\zeta/2}E^{2}(a')}\,, \ee
with $C_1$ a constant. The behavior of Eq.\,(\ref{eqff2}) in the
early epoch, namely when $E(a)\sim (\Omo/\zeta) a^{-3\zeta/2}$, is
$\delta_m(a)\sim a^{3\zeta-2}$. In the case of model I, for which $\zeta=1$, we have $\delta_m(a)\sim a$. This is the same limiting behavior as that of
the $\CC$CDM model, with the proviso that that for both models with $c_0=0$ there is an extra
factor of $\Omo$ in the matter density. Such anomaly is not
innocuous; it has dramatic consequences that will be analyzed in the
next sections.

Before closing this section we should like to point out that the
pure quadratic model $\rL\propto H^2$ (corresponding to
$c_0=\epsilon=0)$ is excluded since such model does not have an
inflection point from deceleration to acceleration (cf. Basilakos,
Polarski \& Sol\`a 2012). In addition, it has no growing modes for
structure formation. This last part can be immediately inferred from
Eq.\,(\ref{eqff2}) using the fact that $E(a)=a^{-3\zeta/2}$ for that
model. As a result, one can easily check that the growing mode
exists only for $\zeta>2/3$ (equivalently, for $\nu<1/3$) and in
this case the Universe is always decelerating. Thus we shall not
consider this model any longer in our analysis. While the pure $H^2$
model is excluded, we should emphasize that when it is complemented
with the $c_0\neq 0$ term, i.e. when we consider $\epsilon=0$ in
type-IV models, the resulting expression takes on the general form
$\rL(H)=C_0+C_2\,H^2$. This structure for the vacuum energy density
is perfectly viable from the phenomenological point of view, and in
fact it is one of the simplest and more attractive formulations of
the dynamical vacuum compatible with the general form of the
effective action in QFT since now both terms (the constant terms and
the $H^2$ term) are allowed by general covariance\,\footnote{For a
theoretical discussion in the context of QFT in curved spacetime,
see e.g. (Sol\`a 2008; Shapiro \& Sol\`a 2009; Sol\`a 2013).}. The
phenomenological status of this model (and some generalizations) has
been confronted against data e.g. in (Basilakos, Plionis \& Sol\`a
2009; Grande, Sol\`a, Basilakos \& Plionis 2011; Basilakos, Polarski
\& Sol\`a 2012) and even more recently in (G\'omez-Valent, Sol\`a \&
Basilakos 2014). It was discussed also in older works both
theoretically and phenomenologically using the first supernovae data
(Espa\~na-Bonet et al. 2003; Shapiro \& Sol\`a 2002, 2003 and 2004).

\section{Vacuum models and linear growth}\label{sect:LinearGrowth}

In Table \ref{tableFit} we show the best-fit values for the models
we are considering. For type-III and type-IV models we have used a
joint statistical analysis involving the latest data, i.e.
SNIa-Union2.1 (Suzuki et al. 2011), BAO measurements in terms of the
parameter $d_z(z_i)=r_s(z_d)/D_V(z_i)$\, (Blake et al. 2011) and the
CMB shift parameter (Ade et al. 2013; Shaefer and Huterer,
2013)\,\footnote{The procedure we have followed is standard, see
e.g. (Basilakos, Plionis \& Sol\`a 2009; Grande, Sol\`a, Basilakos
\& Plionis 2011; G\'omez-Valent, Sol\`a \& Basilakos 2014) for
details.}. We have proceeded in a different manner with type-I and
type-II models due to the fact that the usual fitting formulas for
computing the redshifts at decoupling and the baryon drag epochs
provided by (Hu \& Sugiyama 1995) are tailor-made for the $\CC$CDM
model and in general for $\CC$CDM-like models. While this is the
case for type III and IV models, this is not so for type I and II
for which the additive term is  $c_0= 0$. We have already seen in
Sect. \ref{sect:VacuumModels} that these last two types of models
present some surprises in the structure of the matter density, most
conspicuously the fact that at large redshift they behave
$\rmr\propto\left(\Omo\right)^2$ rather than the standard behavior
$\rmr\propto\Omo$.

For this reason, for the non $\CC$CDM-like models I and II we have
implemented the fitting procedure by just concentrating on the low
and intermediate redshifts, that is to say, we have used the type Ia
supernovae data but avoided using CMB data. At the same time for
these models we have used Eisenstein's BAO parameter $A(z)$
(Eisenstein, 2005), tabulated as in (Blake et al. 2011). It is given
as follows:
\begin{equation}\label{defBAOA}
A({z_i,\bf p})=\frac{\sqrt{\Omo}}{E^{1/3}(z_{i})}
\left[\frac{1}{z_i}\int_{0}^{z_i} \frac{dz}{E(z)} \right]^{2/3}\,.
\end{equation}
For models I and II we have avoided to use the BAO $d_z$-parameter,
which requires the computation of the comoving distance that light
can travel up to the baryon drag epoch (at redshift $z_d$), i.e. the
quantity
\begin{equation}
\label{drag}
r_{s}(z_{d})=\int_{0}^{t(z_d)}\,\frac{c_s\,dt}{a}=\int_{z_d}^{\infty}
\frac{c_s(z)\,dz}{H(z)}\;,
\end{equation}
where
\begin{equation}\label{cs2}
c_s(z)=c\,\left(\frac{\delta {p}_{\gamma}}{\delta
{\rho}_{\gamma}+\delta {\rho}_b}\right)^{1/2}=
\frac{c}{\sqrt{3\,\left(1+{\cal R}(a)\right)}}
\end{equation}
is the sound speed in the baryon-photon plasma. Note that this
quantity is model-dependent because  ${\cal
R}(z)=\delta{\rho}_b/\delta{\rho}_{\gamma}$ is so. However, for
models III and IV (the ones which are $\CC$CDM-like) we can safely
use the BAO $d_z$-parameter, also tabulated in (Blake et al. 2011),
and in fact we have adopted it in such cases. The necessary
corrections for these models amount to the following expression,
which is obtained after using equations (\ref{eq:rhomNR2}) and
(\ref{eq:rhorNR2}):
\begin{equation}\label{eq:RtypeB}
{\cal
R}(t)=\frac{3\Omega_b^{0}}{4\Omega_{\gamma}^{0}}\left[\frac{\sinh\left(\frac{3}{4}\,H_0\,\mathcal{F}\,t\right)}{\sinh\left(\frac{3}{4}\,H_0\,\mathcal{F}\,t_0\right)}\right]^{2/3}.
\end{equation}
One can easily check that for $\nu=\epsilon=0$ we retrieve the
corresponding $\CC$CDM result:
\begin{equation}
\left.{\cal R}(t)\right|_{\epsilon=\nu=0}
=\frac{3\Omega_b^{0}}{4\Omega_{\gamma}^{0}}\,a(t)\,.
\end{equation}
\begin{figure*}
\begin{center}
\includegraphics[scale=0.41]{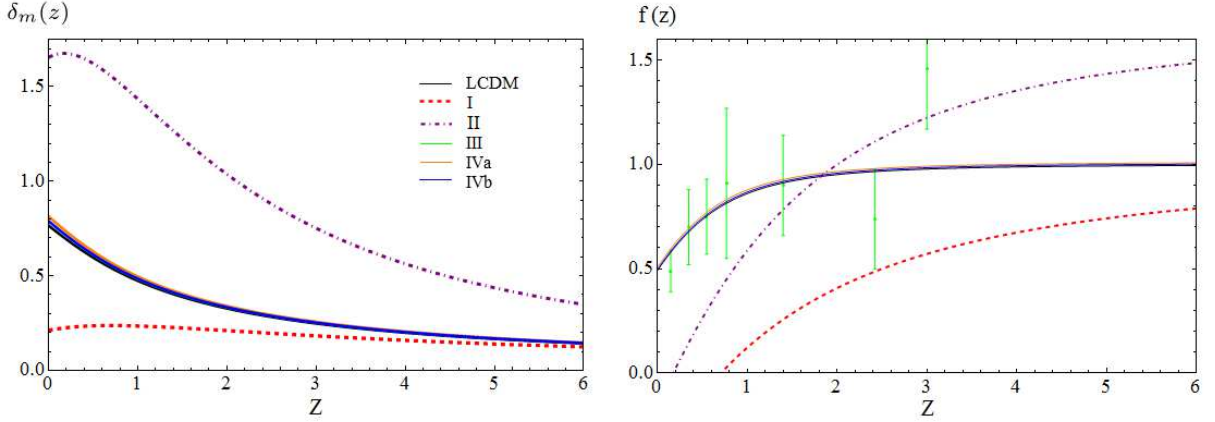}
\caption{\scriptsize {\it Left plot:} The non-normalized density contrast
$\delta_m(z)$ predicted by the various models under study,
Eq.\,(\ref{eq:Models}), using the fit values collected in Table 1;
{\it Right plot:} Comparison of the observational data (see text) --
with error bars depicted in green --  and the theoretical evolution
of the linear growth rate of clustering $f(z)$, confer
Eq.\,(\ref{eq:fz}), for each vacuum model. Models III and IV are
almost indistinguishable from the $\CC$CDM one. \label{fig:f(z)}}
\end{center}
\end{figure*}

As already warned, the situation for models I and II is different as
we cannot use the standard formulae for estimating $z_d$ owing to
the anomalous behavior of $H$ at very high redshift.  For this
reason we have used only the BAO$_A$ data for them (based on the
aforementioned acoustic parameter $A(z)$ whose computation does not
involve any integration in the very high redshift range), and of
course the SNIa data. For models III and IV, in contrast, we have
used SNIa and CMB data collected from the aforementioned references,
and BAO$_{dz}$ data based on the $d_z$-parameter, tabulated also
in\, (Blake et al. 2011).

Proceeding in this way we can see from Table 1 that the fitting
values of $\Omo$ associated to models I and II are not very
different from those of models III and IV, and all of them are
reasonably close to the $\CC$CDM model (which is also included in
that table and fitted from the same data). From this point of view
(and attending also to the $\chi^2$ values per d.o.f.) we can say
that these models perform an acceptable fit to the cosmological
data. For models I and II, however, we can attest this fact only for
the low and intermediate redshift data. If we include the CMB shift
parameter and the BAO$_{dz}$ data, models I and II then peak at
around $\Omo\sim 0.5$ (and with a bad fit quality, see Appendix A).
Such poor performance is caused by the aforementioned
$\rmr\propto\left(\Omo\right)^2$ anomalous behavior of these models
at large redshift.

Even if we restrain to the low and intermediate redshift data for
models I and II, which as we have seen lead to an acceptable value
of $\Omo\simeq 0.3$ (cf. Table 1), they nevertheless clash violently
with a serious difficulty, namely they are bluntly unable to account
for the linear structure formation data, as it is plain at a glance
on Fig.\,\ref{fig:f(z)} (plot on the right). The observational data
in that plot have been taken from Table $1$ of (Jesus et al. 2011)
and references therein.

To better understand the meaning of Fig.\,\ref{fig:f(z)}, let us
recall that from the standard definition of the density contrast
$\delta_m=\delta\rho_m/\rho_m$ one can define the linear growth rate
of clustering (Peebles 1993), as follows:
\be f(z)\equiv\frac{d\,\ln\delta_m}{d\,\ln
a}=-(1+z)\,\frac{d\ln\delta_m(z)}{dz}\,. \label{eq:fz}\ee
Both $\delta_m(z)$ and $f(z)$ have been plotted in
Fig.\,\ref{fig:f(z)} for the models under study together with the
$\CC$CDM.

The obvious departure of models I and II from the linear growth data
is an important drawback for these models. It implies that the
initial success in fitting the Hubble expansion data cannot be
generalized to all low redshift data. Such situation is in contrast
to type III and IV models, which are able to successfully fit the
linear growth data at a similar quality level as the $\CC$CDM, as
can also be appreciated in Fig.\,\ref{fig:f(z)}. In fact, the three
curves corresponding to models III, IV and the $\CC$CDM (for the
best fit values of the parameters in Table 1) lie almost on top of
each other in that figure, whereas the curves for models I and II
depart very openly from the group of $\CC$CDM-like models.  For the
former there is an evident defect of structure formation with
respect to the $\CC$CDM, whilst for the latter there is a notable
excess.

The large differences can be explained as follows. As we have seen
before the ratio $\rho_\CC/\rho_r$ for type-II models is far from 0,
and negative, in the far past. Now, from the basic equations in
Sect. 2 we find that during the matter-dominated epoch the
acceleration of the expansion is given by $\ddot{a}/{a}=(4\pi
G/3)(2\rL-\rmr)$. Thus, a negative value of the vacuum energy
density, $\rL<0$, helps to slow down the expansion (it actually
cooperates with gravitation and enhances the aggregation of matter
into clusters). Actually, the vacuum energy of model II did not
become positive until $H(\tilde{z})=-\epsilon
H_0/\nu\approx 4.13 H_0$, what corresponds to a redshift
$\tilde{z}=3.204$. This is why we obtain larger values of the density
contrast in comparison with the models that take $c_0\ne 0$ (cf.
Fig.\,\ref{fig:f(z)}). Later on the universe
started to speed up, and the transition value from deceleration to acceleration is given by
\be \label{inflectionC1} z_{tr}^{\rm
(II)}=\left[\frac{2(\zeta-\Omo)}{(3\zeta-2)\Omo}\right]^{2/3\zeta}-1\,.
\ee
From the values of the fitted parameters in Table 1, we find $z_{tr}=1.057$. Numerically, it is significantly larger than in the $\CC$CDM ($z_{tr}\simeq 0.69$, for the central fit value of $\Omo$ quoted in Table 1). From this point onwards the type-II vacuum has been accelerating the universe and restraining the gravitational
collapse, but it has left behind a busy history of structure
formation triggered by the large growth rate $\delta_m(a)\sim a^{3\zeta-2}= a^{1.675}$ (cf. the fit value $\zeta=1.225$ from Table 1). Such history is difficult to reconcile with the (much more
moderate) one indicated by observations.

In the other extreme we have type-I model, showing a serious lack of
structure formation as compared to the $\CC$CDM  (cf.
Fig.\,\ref{fig:f(z)}), despite for both models $\delta_m(a)\sim a$.  We can also understand the reason as follows. Let us assume a common value of the density parameter $\Omo$ (which is a good
approximation under the fitting strategy we have followed in Table
\ref{tableFit}). In that case Eq.\,(\ref{eq:rhoLaLinear2}) tells us that the ratio of their vacuum energy
densities is: $\rho^I_\CC/\rho_\CC^{\CC CDM}=1+\Omo(a^{-3/2}-1)$. Thus, during the past cosmic history the vacuum energy density for the type-I model  is positive and always larger
than in the concordance model, so we should expect a reduced growth rate as compared to the $\CC$CDM. This is confirmed in Fig.\,\ref{fig:f(z)}.

In the next section, we analyze the nonlinear perturbation effects
at small scales and consider the different capability of the vacuum
models under study to produce cluster-size halo structures in the
universe. This study will give strength to the results obtained at
the linear level.

%


\section{Number counts analysis}\label{sect:NumberCounts}

In the previous section we have shown that the $\CC$CDM-like vacuum
models III and IV deviate mildly from the concordance model when we
consider the linear structure formation. While in the future it may
be possible to resolve better these differences there is another
useful strategy that can be adopted to magnify the differences to a
larger degree. It is based on the clustering properties of the
nonlinear regime at smaller scales and on counting the number of
formed structures in each vacuum framework. Present X-ray and
Sunyaev-Zeldovich surveys, such as eROSITA (Merloni et al, 2012) and
SPT (Bleem et al, 2014), can be very helpful to test these models.
The method ultimately relies on the Press and Schechter (PSc)
formalism (Press \& Schechter 1974) and generalizations thereof. We
will apply it to the various models under study.

From that formalism  one can predict the abundance of bound
structures that have been formed by gravitational collapse. The
comoving number density of collapsed objects at redshift $z$ within
the mass interval $M$ and $M+dM$ takes on the form
\be
n(M,z)dM=-\frac{\bar{\rho}_m(z)}{M}\frac{ln\,\sigma(M,z)}{dM}f(\sigma;\delta_c)\,,\ee
\noindent where $\bar{\rho}_m$ is the comoving background density
and $f(\sigma;\delta_c)$ is the PSc-function. An important parameter
in it is the collapse density threshold $\delta_c$, which we have
computed numerically for our models in Fig.\,\ref{fig:deltac}. In
the original PSc-form, $f_{\rm PSc}(\sigma;\delta_c)=\sqrt{2/\pi}
(\delta_c/\sigma) \exp(-\delta_c^2/2\sigma^2)$. However, in the
present work we adopt the improved one proposed by (Reed et al.
2007), which depends on several additional parameters. Finally,
$\sigma^2(M,z)$ is the mass variance of the smoothed linear density
field. In Fourier space it is given by:
\be \label{sig88} \sigma^2(M,z)=\frac{D^2(z)}{2\pi^2} \int_0^\infty
k^2 P(k) W^2(kR) dk \,. \ee
\noindent In this expression, $D(z)$ is the linear growth factor of
perturbations, i.e. $D(z)\equiv\delta_m(z)$, which we have computed
before for our models, $P(k)$ is the CDM power-spectrum of the
linear density field and finally we have the smoothing function
$W(kR)=3({\rm
  sin}kR-kR{\rm cos}kR)/(kR)^{3}$, which is the Fourier transform of the following geometric top-hat function with spherical symmetry:  $f_{\rm top\,hat}(r)=3/(4\pi R^3)\,\theta(1-r/R)$, where $\theta$ is the Heaviside function.
It contains on average a mass $M$ within a comoving radius $R=(3M/
4\pi \bar{\rho})^{1/3}$.

\begin{figure*}
\begin{center}
\includegraphics[scale=0.33]{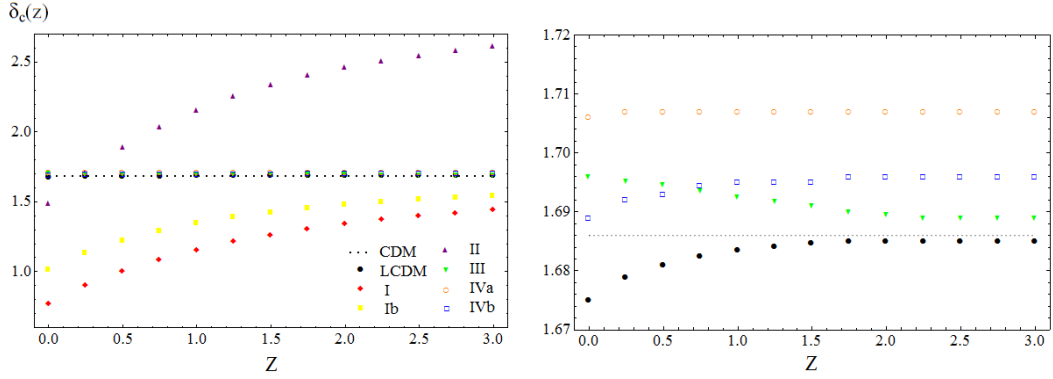}
\caption{\scriptsize Computation of the collapse density threshold $\delta_c(z)$
using the best fit values shown in Table \ref{tableFit}. With these
values we solve numerically Eq.\,(\ref{eq:SDE2}) following  the
procedure outlined in Appendix A of (Grande, Sol\`a, Basilakos \&
Plionis 2011). In both plots we include the fiducial constant CDM
value $\delta_c=\frac{3}{20}(12\pi)^{2/3}\approx 1.686$ (horizontal
dotted line) and the $\CC$CDM curve (solid points, in black). The
models with $c_0\ne 0$ (i.e. III and IV) provide $\delta_c(z)$ very
close to the $\CC$CDM model and the corresponding curves are
cluttered in the plot on the left. In the right plot we zoom in the
relevant region $\delta_c^{CDM}$$^{+0.024}_{-0.016}$ in order to
clearly appreciate the differences between them. In the plot on the
left these differences cannot be seen owing to the large deviations
shown by models I and II ($c_0=0$) which required to use a large
span for the vertical axis. Finally, the curve indicated as Ib has been computed
for model I under another set of inputs (cf. Appendix A of the current paper).
\label{fig:deltac}}
\end{center}
\end{figure*}

The CDM power spectrum
$P(k)=P_{0} k^{n_s} T^{2}(\Omega_{m}^{0},k)$ is used,
%
where $P_0$ is a normalization constant, and $n_s$ is the spectral
index given by $n_s=0.9603\pm 0.0073$ as measured by Planck+WP\,(Ade
et al. 2013). Finally, $T(\Omega_{m}^{0},k)$ is the BBKS transfer
function (Bardeen, Bond, Kaiser \& Szalay 1986; Liddle \& Lyth
2000). Introducing the dimensionless variable $x=k/k_{eq}$, in which
$k_{eq}=a_{eq}H(a_{eq})$ is the value of the wave number at the
equality scale of matter and radiation, we can write the transfer
function as follows:
\begin{eqnarray}\label{jtf}
  T(x) =  \frac{\ln (1+0.171 x)}{0.171\,x}
   \Big[1+0.284 x + (1.18 x)^2 + \, (0.399 x)^3+(0.490
x)^4\Big]^{-1/4}\,.\nonumber
\end{eqnarray}
\noindent It is important to emphasize that $k_{eq}$ is a model
dependent quantity. For type-III and type-IV models one can use the
same formula that is obtained in the $\CC$CDM, due to the fact that
the deviations are negligible in these cases, as we have checked. On
the contrary, with type-I and type-II models we are not allowed to
do that. We must derive the corresponding expression for $k_{eq}$ by
applying (\ref{eq:EaIIrhomrhoR}) and (\ref{eq:aeq}). The final
results for each model read as follows:
\be ({\rm I})\qquad
k_{eq}=H_0\,\sqrt{\frac{2}{\Omega_r^{0}}}\,\left(\Omo\right)^{4/3}\,,\
\ \ \ \ ({\rm II})\qquad
k_{eq}=\frac{H_0\sqrt{2}}{\zeta^{\frac{1}{3\zeta}+\frac{1}{2}}}\left(\Omo\right)^{
2-\frac{2}{3\zeta}}
\left(\Omega_r^0\right)^{\frac{1}{\zeta}-\frac{3}{2}}\,, \ee
and
\be ({\rm III,IV})\qquad
k_{eq}=H_0\,\Omo\,\sqrt{\frac{2}{\Omega^0_r}}\,e^{-\Omega_b^{0}-\sqrt{2h}\frac{\Omega_b^{0}}{\Omo}}\,.
\ee
\begin{figure*}
\begin{center}
\includegraphics[scale=0.33]{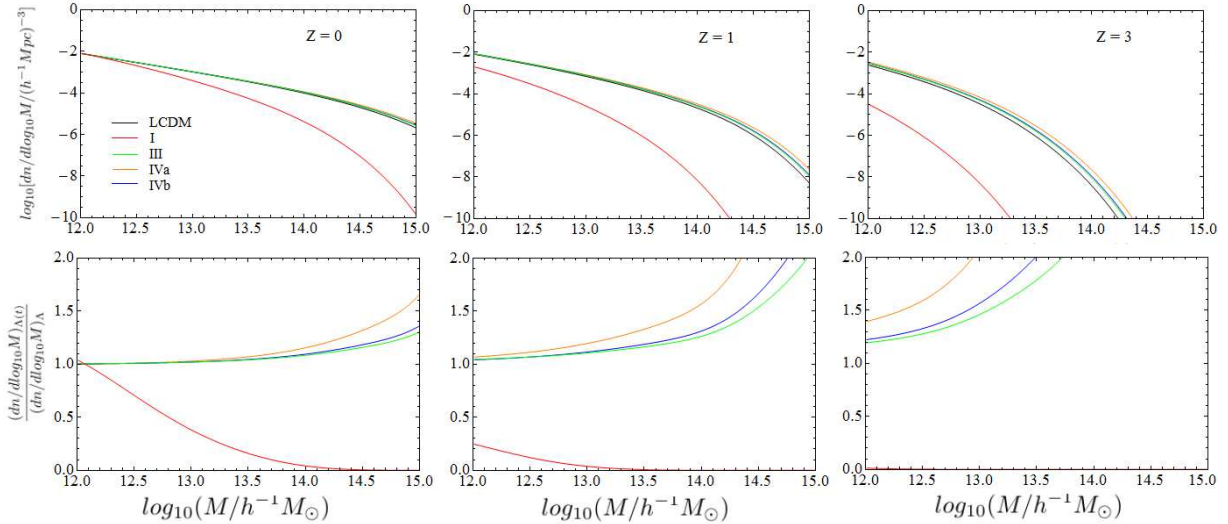}
\caption{\scriptsize {\it Upper plots:} The differential comoving number density
as a function of the halo mass for the various dynamical vacuum
models and the concordance $\Lambda$CDM model at redshifts $z=0$,
$z=1$ and $z=3$, respectively. {\it Lower plots:} Corresponding
differences in the comoving number density with respect to the
$\CC$CDM model.\label{fig:dndlogM}}
\end{center}
\end{figure*}
We normalize the power spectrum using $\sigma_8$, the rms mass
fluctuation amplitude on scales of $R_{8}=8 \; h^{-1}$ Mpc at
redshift $z=0$ [$\sigma_{8} \equiv \sigma_8(0)$].
%
\noindent The $\sigma_8$ value for the different dynamical vacuum
models can be estimated as in (Grande, Sol\`a, Basilakos \& Plionis
2011) by scaling the $\CC$CDM value
%
%
 $\sigma_{8,\Lambda}=0.829\pm 0.012$  extracted from (Ade et al. 2013).
Upon using Eq.\,(\ref{sig88}) with the CDM power spectrum the mass
variance of the linear density field for each model can finally be
computed as follows:
\begin{equation}\label{eq:variance}
\frac{\sigma^2(M,z)}{\sigma^2_{8,\Lambda}}=\frac{D^2(z)}{D^2_{\Lambda}(0)}
\frac{\int_{0}^{\infty} k^{n_s+2} T^{2}(\Omega_{m}^{0}, k) W^2(kR)
dk} {\int_{0}^{\infty} k^{n_s+2} T^{2}(\Omega_{m,\Lambda}^{0}, k)
W^2(kR_{8}) dk}\,.
\end{equation}
Using this procedure, along with the best-fit values of Table
\ref{tableFit} and the numerically determined collapse density
$\delta_c(z)$ (cf. Fig.\,\ref{fig:deltac}) entering the generalized
PSc-function $f(\sigma;\delta_c)$ of (Reed et al. 2007), we have
computed the fractional difference $\delta\mathcal{N}/\mathcal{N}$
(where $\delta\mathcal{N}\equiv\mathcal{N}-\mathcal{N}_{\CC{\rm
CDM}}$) for the number counts of clusters between the dynamical
vacuum models and the concordance $\Lambda$CDM one. The differential
comoving number density of predicted cluster-size structures at
particular values of the redshift ($z=0,z=1$ and $z=3$), as well as
the normalized results with respect to the corresponding $\CC$CDM
prediction, are presented in Fig.\, \ref{fig:dndlogM}, whereas in
Fig.\,\ref{fig:n} we show the differences in the halo mass function
through the comoving number density  for the various models at two
fixed redshifts ($z=1$ and $z=3$).  Finally, in
Fig.\,\ref{fig:DeltaN} we plot the redshift distribution of the
total number of counts.

These figures encapsulate all the main information on the number
counts analysis. They display the number of counts for each
model per mass range at fixed redshift, and the total number of
structures at each redshift within the selected mass range. The
upshot from our analysis is that the models with $c_0\ne 0$ predict
either a very small (type-I) or a very large (type-II) number of
clusters as compared to the $\CC$CDM. This is not surprising if we
inspect the power for structure formation of these models in the
linear perturbation regime (see Fig.\,\ref{fig:f(z)} and the
comments at the end of Sect.\,\ref{sect:LinearGrowth}). As a result
we deem unrealistic the situation for both the type I and type II
models. When we translate this situation to the corresponding
prediction for the number counts we find that, for model I,  ${\cal
N}^{\rm I}/{\cal N}_{\CC{\rm CDM}}\ll 1$,   whereas for model II
${\cal N}^{\rm II}/{\cal N}_{\CC{\rm CDM}}\gg 1$ in the whole range.
As a result, the former yields $\delta{\cal N}/{\cal N}_{\CC{\rm
CDM}}\to -1$ at increasing redshifts (as can be appreciated in
Fig.\,\ref{fig:DeltaN}), whereas the latter is out of the
window under study.

\begin{figure*}
\includegraphics[scale=0.37]{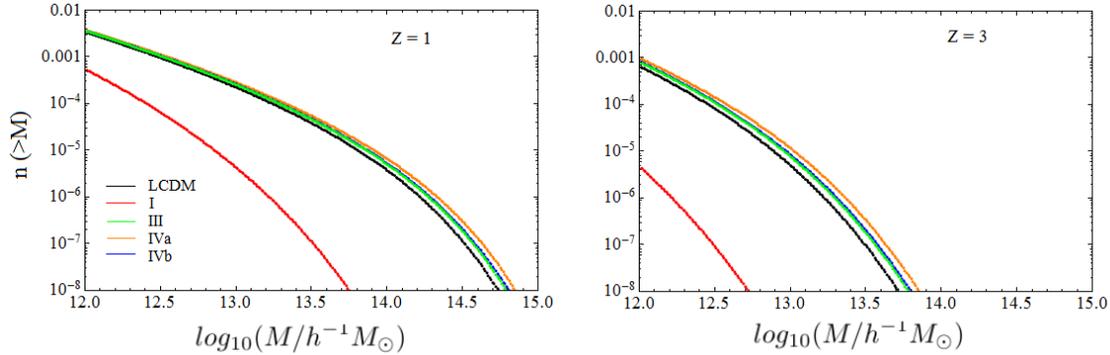}
\caption{\scriptsize The comoving number density at two different redshifts for
the different models.\label{fig:n}}
\end{figure*}
%


In contrast, the situation with the $\CC$CDM-like models III and IV
is quite encouraging. These models represent viable alternatives, at
least from the phenomenological point of view, to the strictly rigid
situation of the $\CC$CDM  (in which $\rL=$const. for the entire
cosmic history).  While these models depart only mildly from the
$\CC$CDM predictions near our time, the differences become sizeable
deep in the past, but still within bound. Concerning the number
counts differences with respect to the concordance model we
recognize from Fig.\,\ref{fig:DeltaN}  significant ($\sim 20-30\%$)
positive departures at moderate redshift ranges, where the total
number of counts is still sizeable. Therefore the predicted
deviations can be measured, in principle, and could be used as an
efficient method to separate models III and IV.


\section{Conclusions}

In this work we have discussed a class of dynamical vacuum models
whose energy density $\rL$ contains a linear and a quadratic term in
the Hubble rate, $H$, i.e. with the general structure: $\rho_\Lambda(H)=C_0+C_1
H+C_2 H^2$. Models in this class having $C_0\neq 0$ have a
well-defined $\CC$CDM limit when the remaining parameters go to
zero. These models are particularly interesting as they can have a
$\CC$CDM-like behavior near our time but their dynamical nature can
help to better explain the past cosmic history. A particular (but qualitatively different)
subclass of dynamical models is those having $C_0=0$; despite they
do not have a $\CC$CDM limit, models of this sort have been
repeatedly invoked in the literature on several accounts. In
particular, the pure linear model $\rL\propto H$ has been proposed
by different authors trying to relate the value of the cosmological
constant with QCD. It is therefore interesting to closely scrutinize
the phenomenological situation of all these models in the light of
the most recent cosmological data.

\begin{figure}
\begin{center}
\includegraphics[scale=0.5]{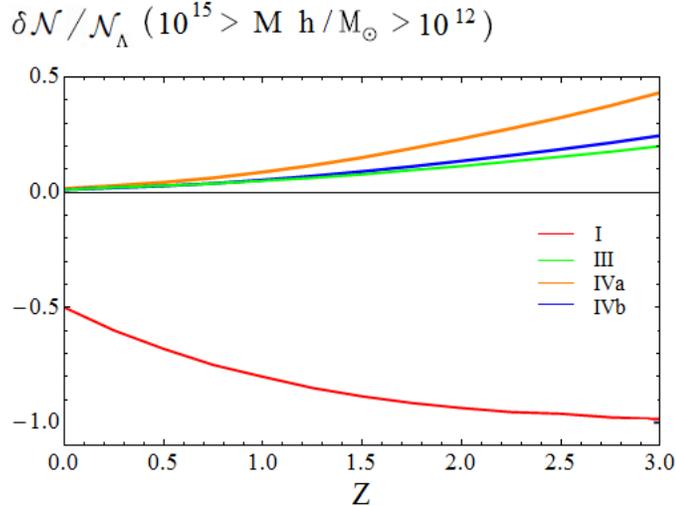}
\caption{\scriptsize The fractional difference
$\delta\mathcal{N}/\mathcal{N}_{\CC{\rm CDM}}$ with respect to the
$\CC$CDM model (where we have defined
$\delta\mathcal{N}\equiv\mathcal{N}-\mathcal{N}_{\CC{\rm CDM}}$).
The curve for the type-II model is not plotted because it is out of
range, i.e. $\delta\mathcal{N}/\mathcal{N}> 1$ .\label{fig:DeltaN}}
\end{center}
\end{figure}

The net outcome of our investigation is the following. At leading
order all these dynamical vacuum models can provide a consistent description of the cosmic evolution, but they
exhibit some differences that can be checked observationally. On a
deeper look, these differences can become quite significant. In particular, we have confronted the vacuum models against the structure formation data, and at the same time we have assessed
their considerably different capability in populating the Universe
with virialized (cluster-size) structures at different redshifts as
compared to the $\CC$CDM model. While all these models can fit
reasonably well the Hubble expansion data, those with $C_0=0$
(denoted as type I and II) are unable to account for the linear
structure formation; and, at the same time, they lead to either an
overproduction or to a drastic depletion in the number of virialized
structures as compared to the $\CC$CDM. In contrast, the $C_0\neq 0$
models (types III and IV) perform  at a comparable level to the
$\CC$CDM and show measurable differences (cf.
Fig.\,\ref{fig:DeltaN}) that could possibly be pinned down in the
near future in ongoing and planned surveys.

The current Universe appears in all these models as FLRW-like,
except that the vacuum energy is not a rigid quantity but a mildly
evolving one. For the $C_0\neq 0$ models the typical values we have
obtained for the coefficients $\nu$ and $\epsilon$ (responsible for
the time evolution of $\rL$) lie in the ballpark of $\sim 10^{-3}$.
This order of magnitude value is roughly consistent with the
theoretical expectations, specially for the coefficient $\nu$ which
can be linked in QFT with the one-loop $\beta$-function of the
running cosmological constant. It is a rewarding feature since it
points to a possible fundamental origin of the structure of these
models in the context of QFT in curved spacetime. However, the
presence of the linear term in $H$ cannot be directly related to a
similar QFT origin, although it could be associated to the presence
of phenomenological bulk viscosity effects. We cannot
exclude this possibility a priori and for this reason we have
performed a thorough phenomenological analysis including this term
in the general structure of the vacuum energy density. Our
conclusion is that the linear term (parameterized by the coefficient
$\epsilon$) is currently tenable at the level $|\epsilon|\sim
10^{-3}$ provided $C_0\neq 0$ (hence for type III and IV models
only). For $C_0=0$, though, the large departure from the $\CC$CDM behavior is unacceptable both within the linear and nonlinear regimes.

To summarize, the wide class of dynamical vacuum models of the cosmic evolution with $C_0\neq0$ may offer an appealing and phenomenologically consistent
perspective for describing dark energy. These models treat the vacuum energy density as a cosmic variable on equal footing to the matter energy density. In a context of an expanding universe this option may be seen as more reasonable than just postulating an everlasting and rigid cosmological
term for the full cosmic history. Some of the models we have
investigated mimic to a large degree the current behavior of the
concordance $\CC$CDM model, but show measurable differences when we
explore our past. Overall the dynamical vacuum models may eventually
offer a clue for a better understanding of the origin of the
$\CC$-term and the cosmological constant problem in the context of
fundamental physics.


\section{Acknowledgements}
The work of AGV has been partially supported by an APIF predoctoral
grant of the Universitat de Barcelona. JS has been supported in part
by FPA2013-46570 (MICINN), Consolider grant CSD2007-00042 (CPAN) and
by 2014-SGR-104 (Generalitat de Catalunya). We thank S. Basilakos
for discussions.


\appendix

\begin{figure}
\begin{center}
\includegraphics[scale=0.4]{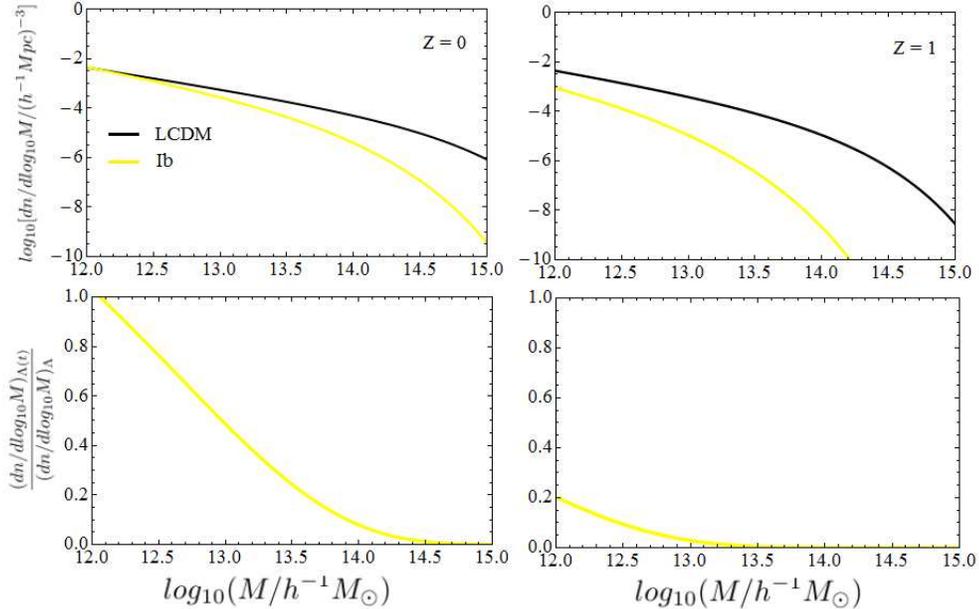}
\caption{\scriptsize As in Fig.3, but for the parameters and the framework used
in (Chandrachani et al. 2014) for $z=0$ and $z=1$. The curve indicated as Ib (showing a negative departure with respect to the $\CC$CDM) corresponds to the new evaluation of model I under the inputs indicated in the text of Appendix A. The corresponding collapse density threshold $\delta_c(z)$ for the new inputs is indicated in Fig. 2 also as Ib.
\label{fig:CompCnro}}
\end{center}
\end{figure}

\section{Number counts for $\rL\propto H$ under different inputs}

In this appendix we briefly compare our results for the linear
model, $\rL\propto H$ -- model I in (\ref{eq:Models}) -- with those
presented in (Chandrachani et al. 2014), where an excess in the
number of counts was reported as compared with the $\CC$CDM. Here we
try to use the parameters indicated by these authors (despite that
not all of them are evident); in particular, we adopt at this point the
halo mass function of (Sheth \& Tormen, 1999). However, after all these changes we do not concur with
their results and we find once more (as in the previous
Fig.\,\ref{fig:dndlogM} for our original fitting parameters, with
the halo mass function of Reed et al., 2007) a large deficit in the
number of counts (cf. Fig.\,\ref{fig:CompCnro}).  Even neglecting
the radiation corrections to the vacuum energy and adopting their
ansatz $k_{eq}\propto\left(\Omo\right)^2$ and the quoted value for
$\Omo=0.45$, we do not meet the claimed excess $\delta{\cal N}>0$
for model I. We also find that by restricting our fit to CMB data
only, the model yields a good quality fit for $\Omo\sim 0.6$, but
only at the expense of a bad fit to SNIa/BAO. If, in addition, we
attempt an overall fit to SNIa+BAO+CMB we find $\Omo\sim 0.52$ with
poor statistical quality ($\chi^2/{\rm d.o.f.}\sim 1.3$).
In short, we find very hard to obtain $\Omo$ near $0.45$ at an acceptable
value of $\chi^2/{\rm d.o.f.}<1$. Even trying to mimic as much as
possible the conditions used by the aforementioned authors we always find, in contrast to them, a large deficit in the number counts (see
Fig.\,\ref{fig:CompCnro}). Our results are consistent with the
rather depleted linear growth behavior exhibited by model I in
Fig.\,\ref{fig:f(z)}, which cannot be reconciled with  $\delta{\cal
N}>0$ neither qualitatively nor quantitatively. Let us also note
that our results for model II (which in this case do predict a large
excess in the number of counts, for the fitted values in Table 1) are also consistent with the large
enhancement of the growth rate displayed by model II in
Fig.\,\ref{fig:f(z)} as compared to the rest of the vacuum models,
including the $\CC$CDM.


\newcommand{\CQG}[3]{{ Class. Quant. Grav. } {\bf #1} (#2) {#3}}
\newcommand{\JCAP}[3]{{ JCAP} {\bf#1} (#2)  {#3}}
\newcommand{\APJ}[3]{{ Astrophys. J. } {\bf #1} (#2)  {#3}}
\newcommand{\AMJ}[3]{{ Astronom. J. } {\bf #1} (#2)  {#3}}
\newcommand{\APP}[3]{{ Astropart. Phys. } {\bf #1} (#2)  {#3}}
\newcommand{\AAP}[3]{{ Astron. Astrophys. } {\bf #1} (#2)  {#3}}
\newcommand{\MNRAS}[3]{{ Mon. Not. Roy. Astron. Soc.} {\bf #1} (#2)  {#3}}
\newcommand{\PR}[3]{{ Phys. Rep. } {\bf #1} (#2)  {#3}}
\newcommand{\RMP}[3]{{ Rev. Mod. Phys. } {\bf #1} (#2)  {#3}}
\newcommand{\JPA}[3]{{ J. Phys. A: Math. Theor.} {\bf #1} (#2)  {#3}}
\newcommand{\ProgS}[3]{{ Prog. Theor. Phys. Supp.} {\bf #1} (#2)  {#3}}
\newcommand{\APJS}[3]{{ Astrophys. J. Supl.} {\bf #1} (#2)  {#3}}

\newcommand{\Prog}[3]{{ Prog. Theor. Phys.} {\bf #1}  (#2) {#3}}
\newcommand{\IJMPA}[3]{{ Int. J. of Mod. Phys. A} {\bf #1}  {(#2)} {#3}}
\newcommand{\IJMPD}[3]{{ Int. J. of Mod. Phys. D} {\bf #1}  {(#2)} {#3}}
\newcommand{\GRG}[3]{{ Gen. Rel. Grav.} {\bf #1}  {(#2)} {#3}}


\end{document}